\documentclass[prl,aps,floatfix,twocolumn,superscriptaddress]{revtex4-1}

\usepackage{amsmath}
\usepackage{ amssymb }
\usepackage{color,soul}
\usepackage{graphicx}
\usepackage{empheq}
\usepackage{verbatim}

\newcommand{\texp}{T_\mathrm{exp}}

\begin{document}

\title{Self-similar structure and experimental signatures of suprathermal ion distribution in inertial confinement fusion implosions}

\author{Grigory~Kagan\footnote{Email: kagan@lanl.gov}}
\affiliation{Los Alamos National Laboratory, Los Alamos, NM 87545}
\author{D. Svyatskiy}
\affiliation{Los Alamos National Laboratory, Los Alamos, NM 87545}
\author{H. G.~Rinderknecht}
\affiliation{Massachusetts Institute of Technology, Cambridge, MA 02139}
\author{M.~J.~Rosenberg}
\affiliation{Massachusetts Institute of Technology, Cambridge, MA 02139}
\affiliation{Laboratory for Laser Energetics, University of Rochester, Rochester, NY 14623}
\author{A. B. Zylstra}
\affiliation{Massachusetts Institute of Technology, Cambridge, MA 02139}
\author{ C.-K. Huang}
\affiliation{Los Alamos National Laboratory, Los Alamos, NM 87545}
\author{ C. J. McDevitt}
\affiliation{Los Alamos National Laboratory, Los Alamos, NM 87545}


\begin{abstract}
The distribution function of suprathermal ions is found to be self-similar under conditions relevant to inertial confinement fusion hot-spots. By utilizing this feature, interference between the hydro-instabilities and kinetic effects is for the first time assessed quantitatively to find  that the instabilities substantially aggravate  the fusion reactivity reduction. The ion tail depletion is also shown to lower the experimentally inferred ion temperature, a novel kinetic effect that may explain the discrepancy between the exploding pusher experiments and rad-hydro simulations and contribute to the observation that temperature inferred from DD reaction products is lower than from DT  at National Ignition Facility.
\end{abstract} 

\maketitle



Recent exploding pusher experiments~\cite{casey-prl, rinderknecht_prl_2014, rosenberg_prl_2014, rosenberg_pop_2014,rinderknecht_pop_2014, rinderknecht_prl_2015} 
 reveal substantial kinetic effects on the implosion performance.  Specific mechanisms potentially responsible for these observations include the inter-ion-species diffusion~\cite{amendt-prl, amendt-pop, electro-diffusion, thermo-diffusion} and reactivity reduction due to ion tail depletion
~\cite{henderson_1974, petschek_1979, molvig_2012, schmit_2013, albright_2013, tang-McDevitt_2014, 
davidovits_2014, cohen_2014}.  
Theoretical evaluation of these phenomena is challenging, however, and while fully kinetic simulations allow study of a certain stage of implosion in specific configurations~\cite{larroche-2012, bellei-2013}, such calculations are computationally prohibitive for modeling of a realistic inertial confinement fusion (ICF) experiment. A substantial simplification results from  treating thermal and suprathermal ions separately~\cite{kagan-APS-2014}. For the former, the mean-free-path $\lambda^{(0)}$ is often much smaller than the characteristic scales of the system $L$, making  fluid equations (including inter-species diffusion) a valid model. The latter constitute only a small fraction of the ion density, momentum, and energy and do not appear explicitly in the fluid equations. However, it is the suprathermal ions that are most likely to undergo fusion reactions, so they do affect the fluid equations implicitly as an energy source.  For these ions, the mean-free-path is much larger than $\lambda^{(0)}$ and can be comparable to $L$ even if $\lambda^{(0)} \ll L$. Hence,  self-consistent modeling of ICF implosions would appear to require a kinetic treatment of suprathermal ions capable of predicting the fusion reactivity at each time step of the fluid equations' evolution.

While suprathermal ions can be described by a reduced linear (as opposite a to fully nonlinear) kinetic equation~\cite{ helander-sigmar}, this task is still non-trivial.  
All prior studies rely on either direct numerical solution~\cite{schmit_2013, tang-McDevitt_2014, davidovits_2014, cohen_2014}   or phenomenological assumptions that affect the structure of the kinetic equation~\cite{ molvig_2012, albright_2013}. Until now, no simple
solution to first-principles kinetic equation for  the suprathermal ions has been found even in the one-dimensional (1D) planar case. The issue becomes particularly pressing in light of hydro-instabilities at the fuel-pusher interface~\cite{chikang-prl-2002,regan-prl-2013, smalyuk-prl-2014, thomas-prl-2012, aglitskiy-prl-2001, velikovich-pop-2001,radha-pop-2005, clark-pop-2013}. It is near this interface that the suprathermal ion distribution is modified  most, so one should expect substantial interference between the instabilities and the fusion reactivity. However, applying direct numeric methods to a complicated geometry is quite difficult and quantitative assessment of this interference has not been presented.  

In this Letter we demonstrate a physically intuitive, semi-analytical solution  to the first-principles kinetic equation for suprathermal ions. 
This results from the self-similar structure of the ion distribution, scaling with the distance to the interface relative to the square of the ion energy. In the 1D planar geometry, the solution  agrees precisely with direct numerical results. Furthermore, comparison with the numeric solution for the 1D spherical geometry shows that the self-similar structure is robust against perturbations of the interface from the planar geometry. This  allows us for the first time to evaluate the impact of hydro-instability on the reactivity reduction, which is  found to be substantially enhanced. We also obtain a novel kinetic prediction for ICF experiments: that ion tail depletion results in the experimentally inferred temperature being lower than the actual one.

We consider a spherically symmetric hot-spot with the radius $R_h$ surrounded by a cold pusher. From symmetry, the distribution function $ f_{\alpha}$ of ion species $\alpha$ depends only on the three variables: radial coordinate $r$, particle 
speed $v$, and pitch angle $\theta$ between velocity and radius vectors. Defining   $\mu \equiv \cos{\theta}$, we obtain the time-stationary Vlasov operator 
\begin{equation}
\label{eq: vlasov}
 \vec{v}\cdot\nabla f_{\alpha} =  v\Bigl[\mu \frac{\partial f_{\alpha}}{\partial r} + \frac{(1-\mu^2)}{r} \frac{\partial f_{\alpha}}{\partial  \mu} \Bigr].
 \end{equation}

The collision operator for species $\alpha$ is  
$C_{\alpha}\left\{f_{\alpha} \right\} = \sum_{\beta} C_{\alpha\beta}\left\{f_{\alpha}\right\}$,
where $C_{\alpha\beta}$ denotes collisions of  ion species $\alpha$ with  ion species $\beta$. We neglect  ion-electron collisions 
%
and note that the suprathermal ions mostly collide with thermal ions, which are close to Maxwellian everywhere outside a narrow vicinity of the boundary. Therefore, for suprathermal ions of species $\alpha$,  \cite{ helander-sigmar, NRL} 
\begin{eqnarray}
\nonumber
C_{\alpha\beta}\left\{f_\alpha \right\} \approx 
 \nu_{\alpha\beta}  \frac{v_{T\alpha}^3}{2 v^3} \frac{\partial }{\partial \mu} (1-\mu^2)\frac{\partial f_\alpha}{\partial \mu} + \\
 \label{eq: coll-op-2}
 \nu_{\alpha\beta} \frac{m_\alpha}{m_\beta} \frac{v_{T\alpha}^3}{v^2}\frac{\partial}{\partial v} \Bigl(f_\alpha+ \frac{T_\beta}{m_\alpha}\frac{1}{v}\frac{\partial f_\alpha}{\partial v}          \Bigr),
\end{eqnarray}
where $m_{\alpha}$,  $T_{\alpha}$ and $v_{T\alpha} \equiv \sqrt{2 T_{\alpha}/m_{\alpha} }$ denote the particle mass, bulk temperature and thermal velocity of species $\alpha$, respectively, and the collision frequency is defined by
\begin{equation}
\label{eq: coll-freq}
\nu_{\alpha\beta} =  \frac{4\pi n_{\beta} Z_{\alpha}^2 Z_{\beta}^2 e^4 \ln{\Lambda}}{m_{\alpha}^2 v_{T\alpha}^3},
\end{equation}
where $Z_{\alpha}$ and $n_{\alpha}$ are the charge number and the bulk ion density of species $\alpha$, respectively, and $\ln{\Lambda}$ is the Coulomb logarithm.

The rate of  energy exchange between  thermal ions of species $\alpha$ and $\beta$ is on the order of $\nu_{\alpha\beta} \sim \nu_{\alpha\alpha} \sim \nu_{\beta\beta} $, making their bulk temperatures equal, i.e. $T_{\alpha} \equiv T_0$ for all $\alpha$. Assuming a 
flat temperature profile in the hot spot and equating the right sides of Eqs.~(\ref{eq: vlasov}) and~(\ref{eq: coll-op-2})  yield the stationary kinetic equation for the tail of $f_\alpha$
\begin{eqnarray}
\nonumber
\mu \frac{\partial f_{\alpha}}{\partial x} +  \frac{1- \mu^2}{x}  \frac{\partial f_{\alpha}}{\partial \mu} = \\
\label{eq: kin-eq-1}
\frac{1}{N_K^{(\alpha)}}  \Bigl[ 
\frac{ 1}{2 \varepsilon^{2}}\frac{\partial }{\partial \mu} (1-\mu^2)\frac{\partial f_{\alpha}}{\partial \mu}
+
\frac{2 G_{\alpha}}{ \varepsilon} \frac{\partial }{\partial \varepsilon} \Bigl( f_{\alpha} + \frac{\partial f_{\alpha}}{\partial \varepsilon}          \Bigr)  \Bigr],
\end{eqnarray}
where $x \equiv r/R_h$, $\varepsilon \equiv m_{\alpha} v^2/2T_0 $, $N_K^{(\alpha)} \equiv v_{T\alpha}/(R_h \sum_{\beta} \nu_{\alpha\beta})$, and
$G_{\alpha} \equiv \left( \sum_{\beta} (m_{\alpha}/m_{\beta}) n_{\beta} Z_{\beta}^2 \right)    \big/ \left( {\sum_{\beta} n_{\beta} Z_{\beta}^2} \right) $, 
%
with summations over all ion species including $\alpha$.  The parameter $G_{\alpha}$ depends only on the relative concentrations of the bulk ion densities; for example, in a 50/50 DT mixture, $G_D = 5/6$ and $G_T = 5/4$. The Knudsen number  $N_K^{(\alpha)}$ is the ratio of the effective mean free path of a thermal ion 
$\lambda_{\alpha}^{(0)} \equiv v_{T\alpha}/( \sum_{\beta} \nu_{\alpha\beta})$ and $R_h$ and thus is the key parameter quantifying importance of the kinetic effects. 

The kinetic equation~(\ref{eq: kin-eq-1})  needs to be accompanied by a condition at the interface between the hot and cold plasmas. A natural constraint results from assuming that there is no suprathermal ion inflow from the pusher into the hot-spot
\begin{equation}
\label{eq: constraint-1}
f_{\alpha}(x=1,-1\le \mu \le 0, \varepsilon) =0.
\end{equation}
In addition, the distribution function must be isotropic at the center due to symmetry
\begin{equation}
\label{eq: constraint-2}
\partial f_{\alpha}(x=0,\mu , \varepsilon)/\partial \mu =0.
\end{equation}
Finally, inside the hot-spot we expect $f_{\alpha}$ to become Maxwellian as $\varepsilon$ approaches $1$ from above since thermal ions are assumed to be close to equilibrium. 

Physically, one expects the solution to be effectively planar when the mean free path of a suprathermal ion with energy $\varepsilon$, 
$\lambda_{\alpha}^{(\varepsilon)} \equiv \varepsilon^2 \lambda_{\alpha}^{(0)}$, is much less than $R_h$, or 
\begin{equation}
\label{eq: planar-limit}
N_K^{(\alpha)}  \varepsilon^2 \ll 1.
\end{equation}
From direct numerical solution to the PDE problem formulated by Eq.~(\ref{eq: kin-eq-1}) along with constraints~(\ref{eq: constraint-1}) and (\ref{eq: constraint-2}) one can find that in this limit the distribution is self-similar, 
$f_{\alpha}(x,\mu, \varepsilon) =f_M \phi(\frac{1-x}{N_K^{(\alpha) } \varepsilon^2 }  ,\mu)$, where  $f_M = n_{\alpha}(m_{\alpha}/2\pi T_0)^{3/2}\mathrm{e}^{-\varepsilon}$ is Maxwellian.
 This finding has a transparent physical interpretation: for a given $\varepsilon$ deviation from equilibrium  is controlled by the distance to the boundary $y = R_h - r$ normalized to 
 $\lambda_{\alpha}^{(\varepsilon)}$.
We then construct a  solution that has this feature manifestly while keeping the pitch-angle scattering structure of the collision operator
\begin{equation}
\label{eq: gen-sol-2}
f_{\alpha}  = f_M [1 +  \sum_{n} c_{n}  \psi_n (\mu) \mathrm{e}^{- \sigma_n z} ],
\end{equation}
where $z \equiv y/\lambda_{\alpha}^{(\varepsilon)} = y/(\lambda_{\alpha}^{(0)}  \varepsilon^2)$,  
$c_n$ are free constants,
 and eigenvalues and eigenvectors $ \sigma_n$ and $ \psi_n (\mu)$ satisfy
\begin{equation}
\label{eq: eig-mu}
 \sigma_n \psi_n = 
\frac{ 1}{2 \mu}\frac{d}{d \mu} (1-\mu^2)\frac{d  \psi_n}{d \mu} - \frac{3G_{\alpha}}{\mu} \psi_n.
\end{equation}
Validity of this choice will be verified by comparing  the resulting semi-analytical solution against the direct numerical one. We now proceed to identifying $c_n$ that are compatible with conditions~(\ref{eq: constraint-1}) and (\ref{eq: constraint-2}).

The spectrum of the operator of Eq.~(\ref{eq: eig-mu}) is symmetric about  zero (i.e. the eigenvalues come in pairs $\sigma_n $ and $-\sigma_n $), a consequence of the right-left symmetry of the planar case. If the  hot plasma occupies the half space $-\infty < y < 0$,  the constraint~(\ref{eq: constraint-2}) dictates that only eigenfunctions with $\sigma_n < 0$ are included in the expansion~(\ref{eq: gen-sol-2}). Imposing 
constraint~(\ref{eq: constraint-1}) is a more non-trivial task, since it applies to $\mu <0$ only. 

To implement this condition, we evaluate the matrix of the operator~(\ref{eq: eig-mu}) over Legendre polynomials $P_k(\mu)$. Solving the resulting eigenvalue problem gives 
$\psi_n(\mu) = \sum_k a_{nk} P_k(\mu)$. Defining $b_k = 1 + \sum_n c_n a_{nk} $ for $k = 0$ and $b_k =  \sum_n c_n a_{nk} $ otherwise, condition~(\ref{eq: constraint-1}) establishes a matrix relation between the vectors of odd and even coefficients $b_k$
\begin{equation}
\label{eq: constraint-leg-2}
b_{2k+1}  = \sum_{m} D_{mk} b_{2m},
\end{equation}
where $D_{nk} =  (4 k + 3) \chi_{2k, 2n+1}$ with \cite{byerly-1959}
$$ 
\chi_{i, j} = \frac{(-1)^{(i+j+1)/2} i! j!}{2^{(i+j-1)}  (i-j) (i+j+1) [(i/2)!]^2 \{ [(j-1)/2]! \}^2}.
$$ 
Importantly, if $N$ Legendre polynomials are kept in the expansion, this relation is equivalent to  $N/2$ scalar equations, which is exactly equal to the number of unknowns in Eq.~(\ref{eq: gen-sol-2}) after eliminating modes with positive eigenvalues \cite{exist-fnote}. 
Once the eigenvectors $a_{nk}$ are calculated, Eq.~(\ref{eq: constraint-leg-2}) gives $c_n$ and therefore the distribution function through Eq.~(\ref{eq: gen-sol-2}). 
\begin{figure}[htbp]
\includegraphics[width=0.4\textwidth,keepaspectratio]{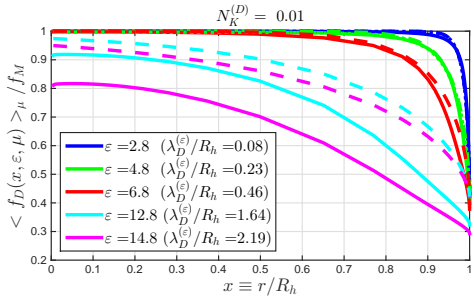}
\caption{Spatial dependence of the pitch-angle averaged distribution function relative to Maxwellian for several values of energy, as obtained from the direct numerical solution of Eq.~(\ref{eq: kin-eq-1}) (solid) and the semi-analytical  solution (dashed).}
\label{fig: n_of_x}
\end{figure}

To verify that this solution is  precise in the planar case and investigate its robustness against deviations of the interface from the planar geometry, we compare it with the distribution function obtained by direct numerical solution of Eq.~(\ref{eq: kin-eq-1}). Fig~\ref{fig: n_of_x} shows a 
comparison of the deuteron distribution function relative to Maxwellian in the 50/50 DT mixture with $N_K^{(D)}=0.01$ where solid curves are 
to the numerical solution and dashed lines are to the planar solution~(\ref{eq: gen-sol-2}) with $(1-x)/(N_K\varepsilon^2)$ substituted for $z$. For  energies such that $N_K \varepsilon^2 \ll 1$, the  solutions agree very well, which is  consistent with formal planar limit condition~(\ref{eq: planar-limit}). 
Furthermore, 
the agreement is  good even for $N_K \varepsilon^2 \sim 0.5$ (red curve). Comparisons for various $N_K^{(\alpha)}$ and mixture compositions show similar agreement.

When $\lambda_{\alpha}^{(\varepsilon)} \equiv \varepsilon^2 \lambda_{\alpha}^{(0)} > R_h$, 
 the planar solution overestimates the suprathermal ion population, reflecting a shortcoming of the planar model in the context of spherical geometry when ions ``feel" not only the distance to the boundary, but also the distance to the center of the hot-spot. To investigate significance  of this spherical effect we consider the reactivity $\left<\sigma v\right>$.
 
\begin{figure}[htbp]
\includegraphics[width=0.4\textwidth,keepaspectratio]{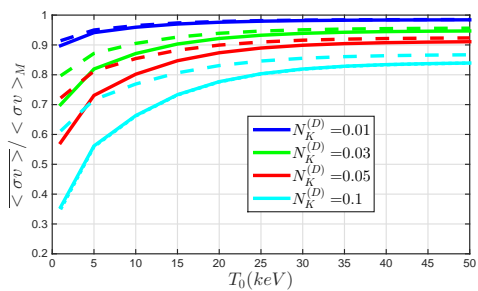}
\caption{Volume averaged DT reactivity relative to Maxwellian computed from  the direct numerical solution of Eq.~(\ref{eq: kin-eq-1}) (solid) and from the semi-analytical  solution (dashed).}
\label{fig: sigma_of_T}
\end{figure}
The fusion cross-section $\sigma$ is defined in absolute energy units, while $f_{\alpha}$ deviation from Maxwellian (as well as deviation between  planar and spherical solutions) is 
governed by the energy $\varepsilon$ normalized to the main ion temperature. Thus, for a fixed $N_K^{(\alpha)}$ and higher  $T_0$ the reactivity is due to lower $\varepsilon$, thereby diminishing  these deviations. Conversely, for fixed $T_0$, larger $N_K^{(\alpha)}$ corresponds to a larger mean-free-path for any given $\varepsilon$, making the effect stronger. This qualitative picture is supported by the results in Fig~\ref{fig: sigma_of_T}, which presents the volume averaged reactivity 
$\overline{\left<\sigma v\right> } \equiv V^{-1} \int_{V} d^3r \left<\sigma v\right>$
for the DT fusion reaction in the 50/50 DT mixture. For $T_0 > 5$~keV,   the predictions of the planar model and the direct numerical calculation are within $15\%$ even for $N_K^{(D)} = 0.1$ and  within $5\%$ for $N_K^{(D)} < 0.03$. 
In the 1-5 keV range the discrepancy is larger, but still within $12\%$ for $N_K^{(D)} = 0.01$ and $0.03$, for which deviation from 1D case will be considered.
Hence, with respect to reactivity, the distance to the boundary is the only relevant scale and the resulting planar solution is robust to perturbations of the boundary from the planar geometry. 

The above  allows us for the first time to  perform a quantitative assessment of  kinetic effects on the reactivity in the presence of hydro-instability. The entire implosion analysis would include modeling the instability along with the kinetic effects. For the purpose of demonstration, here we consider a sample instability resulting in the interface perturbed as $R = R_h + \Delta R \cos {(m \vartheta)}$,
where $R$ and $\vartheta$ are the radial and polar angle coordinates of the hot-spot boundary and symmetry in the azimuthal angle $\varphi$ is assumed. Experiments and simulations indicate that $\Delta R/R$ can be as large as $1/3$  \cite{radha-pop-2005, chikang-prl-2002, clark-pop-2013, smalyuk-prl-2014} and to imitate the spatial structure of the unstable region we take  $m = 20$. Then, we evaluate the overall reactivity reduction as follows: for a point A inside the hot-spot the point B on the perturbed surface is found such that the distance AB is the shortest; the distribution function is evaluated from our solution with AB inserted for $y$;  local reactivity reduction at each point A is calculated and averaged  over the perturbed volume $V$. This procedure is computationally inexpensive and can be  applied readily to an arbitrarily perturbed surface.
\begin{figure}[htbp]
\includegraphics[width=0.4\textwidth,keepaspectratio]{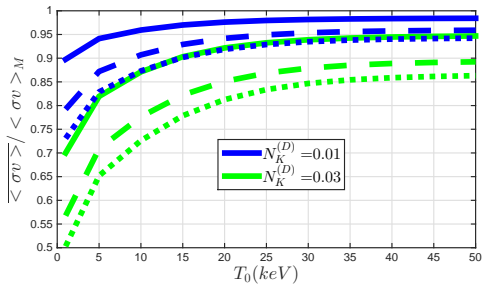}
\caption{Volume averaged DT reactivity relative to Maxwellian computed for the unperturbed spherical hot-spot boundary (solid) and for the boundary perturbed with 
$\Delta R/R_h = 1/5$ (dashed) and $\Delta R/R_h = 1/3$ (dotted).}
\label{fig: reactivity_curved}
\end{figure}
Fig~\ref{fig: reactivity_curved} presents results of this calculation for the DT reaction in a 50/50 DT mixture for $\Delta R/R_h$ equal to $1/5$ (dashed line) and $1/3$ (dotted line) and Knudsen numbers of $0.01$ and $0.03$. Introduction of a perturbation even at the conservative level of $\Delta R/R_h=1/5$ can double the reactivity reduction in the 1-10~keV range. Importantly, the change in reactivity from the surface perturbation is substantially larger than discrepancy 
 between predictions of  the numerical  and semi-analytical  solutions in Fig~\ref{fig: sigma_of_T} for any given $T_0$.
  
\begin{figure}[htbp]
\includegraphics[width=0.4\textwidth,keepaspectratio]{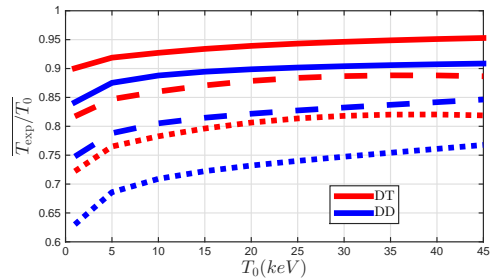}
\caption{Volume averaged DT and DD burn temperature relative to the bulk ion temperature $T_0$ computed from the direct numerical solution for $N_K^{(D)} = 0.05$ (solid), $N_K^{(D)} = 0.1$  (dashed) and  $N_K^{(D)} = 0.2$ (dotted).}
\label{fig: temp_DD_in_DT}
\end{figure}

Finally, we demonstrate a novel effect of the ion tail  depletion on the experimentally inferred ion temperature. In ICF experiments the hot-spot ion temperature is deduced from the width of the reaction products' spectra, which can be related to the mean square of the center-of-mass velocity of ion pairs undergoing this reaction \cite{brysk-1973}
\begin{equation}
\label{eq: T-exp}
\texp = \frac{m_1+m_2}{3}\frac{\int d^3v_1d^3v_2 f_1 f_2 \sigma v u^2_{c.m.}}{\int d^3v_1d^3v_2 f_1 f_2 \sigma v},
\end{equation}
where $v \equiv |\vec{v}| = |\vec{v}_1 -\vec{v}_2|$ and $\vec{u}_{c.m.} = (m_1\vec{v}_1 + m_2 \vec{v}_2)/(m_1+m_2)$. Since the reaction cross section $\sigma$ depends only on $v$, the expression on the right side of  
Eq.~(\ref{eq: T-exp}) can be shown to recover $T_0$ when $f_{1,2}$ are Maxwellian regardless of the fusion reaction. For distribution functions different from $f_M$, one should expect $\texp$ to also be different from $T_0$ and, equally importantly, to depend on 
$\sigma$. The DD cross-section is due to larger energies than the DT cross-section and, in the 50/50 DT mixture, $f_D$ is farther from equilibrium than $f_T$ due to $G_D < G_T$. As a result, $\texp$ associated with DD reaction is lower than that associated with DT reaction, which is confirmed by Fig~\ref{fig: temp_DD_in_DT} showing the two temperatures evaluated from the direct numerical solution according to Eq.~(\ref{eq: T-exp}).
We also see that the reduction in $\texp$ is less than that in $\left<\sigma v \right>$. In addition, it can be found that,   unlike $\left<\sigma v \right>$ case,  it is  the spherical effects on the distribution function rather than its planar structure that reflect stronger on $\texp$. This difference between the $\left<\sigma v \right>$ and $\texp$, to leading order, results from the former being governed by the ion number density within the Gamow window, whereas the latter is 
governed by higher moments and finer features of the distribution function. 
The size of the predicted effect is consistent with the discrepancy between the temperature observed in exploding pusher experiments, in which $N_K$ can be even higher than shown in Fig~\ref{fig: temp_DD_in_DT}, and that predicted by simulations~\cite{rosenberg_prl_2014, rosenberg_pop_2014, rinderknecht_pop_2014, rinderknecht_prl_2015}. 
Of course, in these experiments kinetic effects can also lower the actual temperature by reducing the shock heating; however, employing standard formulae for the spectrum width would diminish the inferred temperature even further.

The DD temperature lower than DT by about 25\% is often observed in cryo-implosions at NIF~\cite{NIF-team}. 
The only explanation available to date is based on the bulk fluid motion \cite{appelbe-ppcf-2011} and a detailed analysis of this effect in application to turbulent implosions is presented in Ref.~\cite{murphy-pop-2014}.
However, to give the 25\% difference between the burn temperatures this mechanism requires all the plasma energy to be kinetic that is hardly possible 
during ICF burn. In a more realistic scenario with the turbulent energy on the order of the thermal energy it can give 10-15\% only~\cite{murphy-pop-2014}. The following estimate demonstrates that the newly predicted effect can be responsible for the remaining part of the difference. 

The turbulence strongly distorts the mixing layer topology, so the approach employed for evaluating the role of instabilities no longer applies. To estimate the above kinetic effects in this situation one can view the mix layer as a suspension of hot plasma droplets in a cold pusher material~\cite{tang-McDevitt_2014}. The effective Knudsen number associated with a given droplet is $\lambda_{\alpha}^{(0)} / L$, where $L$ is the characteristic radius of the droplet  rather than of the hot-spot. The \emph{nominal} Knudsen number for NIF,   $\lambda_{\alpha}^{(0)} / R_h$,  is a few percent~\cite{atzeni-book} and $R_h \gg L$, so the effective Knudsen number should be taken to be a fraction of unity. Results of 
Fig~\ref{fig: temp_DD_in_DT} give that the difference between the DD and DT temperatures grows with $N_K$, becoming about $12\%$ in the 1-5 keV range for $N_K = 0.2$. For higher Knudsen numbers, which are likely according to the above, our approach to suprathermal ions is complicated; yet, the trend suggests that the DD and DT temperatures fall apart even further. Also, the mix layer occupies the outer radii and therefore constitutes a large fraction of the hot-spot. Hence, 10-15\% of the difference between the DD and DT temperatures can be realistically attributed to the ion tail depletion and the newly presented mechanism together with the bulk fluid motion effect considered earlier~\cite{murphy-pop-2014} can fully explain the observations. 

Our predictions are limited by shortcomings of the kinetic equation~(\ref{eq: kin-eq-1}), which assumes a constant ion temperature in the hot-spot. To investigate robustness of the newly found effects with respect to a more realistic temperature profile, we have conducted direct numerical simulations for an isobaric hot-spot with the main ion temperature given by commonly observed $T=T_0 [1-(r/R_h)^2]^{2/7}$ \cite{lindl-book}. The results for the reactivity reduction and the experimentally inferred ion temperatures  in terms of the volume averaged Knudsen number $\overline{N_K^{(D)}}$ and bulk ion temperature $\overline{T}$ turn out to be in reasonable agreement with the results of Figs~\ref{fig: sigma_of_T} and \ref{fig: temp_DD_in_DT} and show the same general trends with these parameters. Furthermore, while the flat profile model presented in this Letter slightly overpredicts the reactivity reduction, it underpredicts the difference between the experimentally inferred DD and DT temperatures. We thus expect that our conclusions qualitatively persist in practical hot-spot configurations, though some quantitative changes are possible.

To summarize, a semi-analytical solution to first-principles equation for the suprathermal ions in 1D geometry has been constructed and provides a computationally expedient tool for investigating kinetic effects in complicated geometries. In particular, the analysis demonstrates that hydrodynamic instabilities at  hot-spot/pusher interfaces can substantially aggravate the reactivity reduction. Moreover, the ion tail depletion results in the experimentally inferred core ion temperatures being lower than the actual ones, which may explain recent measurements in exploding pusher implosions and 
contribute to the observation that DD burn temperature is lower than DT burn temperature at NIF.

\begin{acknowledgments}

The authors would like
to acknowledge  useful conversations with B.J. Albright, K. Molvig, T.J. Murphy, N.M. Hoffman, R.C. Shah, A.N. Simakov, Y.-H. Kim, M.J. Schmitt and H.W. Herrmann of LANL and H. Sio, M. Gatu Johnson, J.A. Frenje, F.H. S\'{e}guin, C.K. Li and R.D. Petrasso of MIT. This work is performed under the auspices of the U.S. Dept. of Energy by the Los Alamos National Security, LLC, Los Alamos National Laboratory under Contract No. DE-AC52-06NA25396. 

\end{acknowledgments}


\begin{thebibliography}{10}
\bibitem{exist-fnote} Unlike the fully kinetic approach, quasi-stationary solution for suprathermal ions exists in the absence of sources since the ion flux leaving through the hot-spot boundary in the physical space is balanced by the flux incoming through the low-energy cutoff.
\end{thebibliography}
\end{document}